\documentclass{article}
\usepackage{spconf,amsmath,graphicx}
\usepackage{amssymb}
\usepackage{booktabs}
\usepackage{multicol}
\usepackage{multirow}
\usepackage{makecell}
\usepackage{hyperref}
\usepackage{enumitem}
\setlist{nosep, leftmargin=14pt}

\usepackage{mwe} 

\title{Self Pre-training with Masked Autoencoders for Medical Image Classification and Segmentation}
\name{%
\begin{tabular}{@{}c@{}}
Lei Zhou$^{\star}$ \qquad Huidong Liu$^{\star\P}$ \qquad Joseph Bae$^{\dagger}$ \qquad Junjun He$^{\ddagger}$ \\ \qquad Dimitris Samaras$^{\star}$ \qquad Prateek Prasanna$^{\dagger}$
\end{tabular}}

\address{$^{\star}$Department of Computer Science, Stony Brook Univeristy, NY, USA\\
$^{\dagger}$Department of Biomedical Informatics, Stony Brook University, NY, USA\\
$^{\P}$ Amazon \qquad $^{\ddagger}$Shanghai Artificial Intelligence Laboratory}

\begin{document}
%
\maketitle
\begin{abstract}
Masked Autoencoder (MAE) has recently been shown to be effective in pre-training Vision Transformers (ViT) for natural image analysis. By reconstructing full images from partially masked inputs, a ViT encoder aggregates contextual information to infer masked image regions. We believe that this context aggregation ability is particularly essential to the medical image domain where each anatomical structure is functionally and mechanically connected to other structures and regions.
Because there is no ImageNet-scale medical image dataset for pre-training,
we investigate a self pre-training paradigm with MAE for medical image analysis tasks. 
Our method pre-trains a ViT on the training set of the target data instead of another dataset. Thus, self pre-training can benefit more scenarios where pre-training data is hard to acquire.
Our experimental results show that MAE self pre-training markedly improves diverse medical image tasks including chest X-ray disease classification, abdominal CT multi-organ segmentation, and MRI brain tumor segmentation. Code is available at \url{https://github.com/cvlab-stonybrook/SelfMedMAE}
\end{abstract}

\section{Introduction}
Within a medical image, each anatomical structure functionally and mechanically  interacts with other structures and regions present in the human body. Image analysis must therefore accounts for these interdependencies and relationships. For instance, classification of pathology on chest x-ray might rely not only on textural changes within lung regions but also on relative changes in cardiac and mediastinum 
\cite{ekeh2008chest}. For segmentation, features inherent to both the target object and surrounding tissue enable delineation of specific structures. The presence of a brain tumor commonly results in additional changes to the tumor's surrounding microenvironment including edema, structural shifts in brain tissue, and increased vascularization \cite{wick2004brain}. 
We hypothesize that enforcing a strict requirement on contextual information learning can improve deep learning--based  medical image analysis.

\begin{figure}[t]
    \centering
    \includegraphics[width=0.47\textwidth]{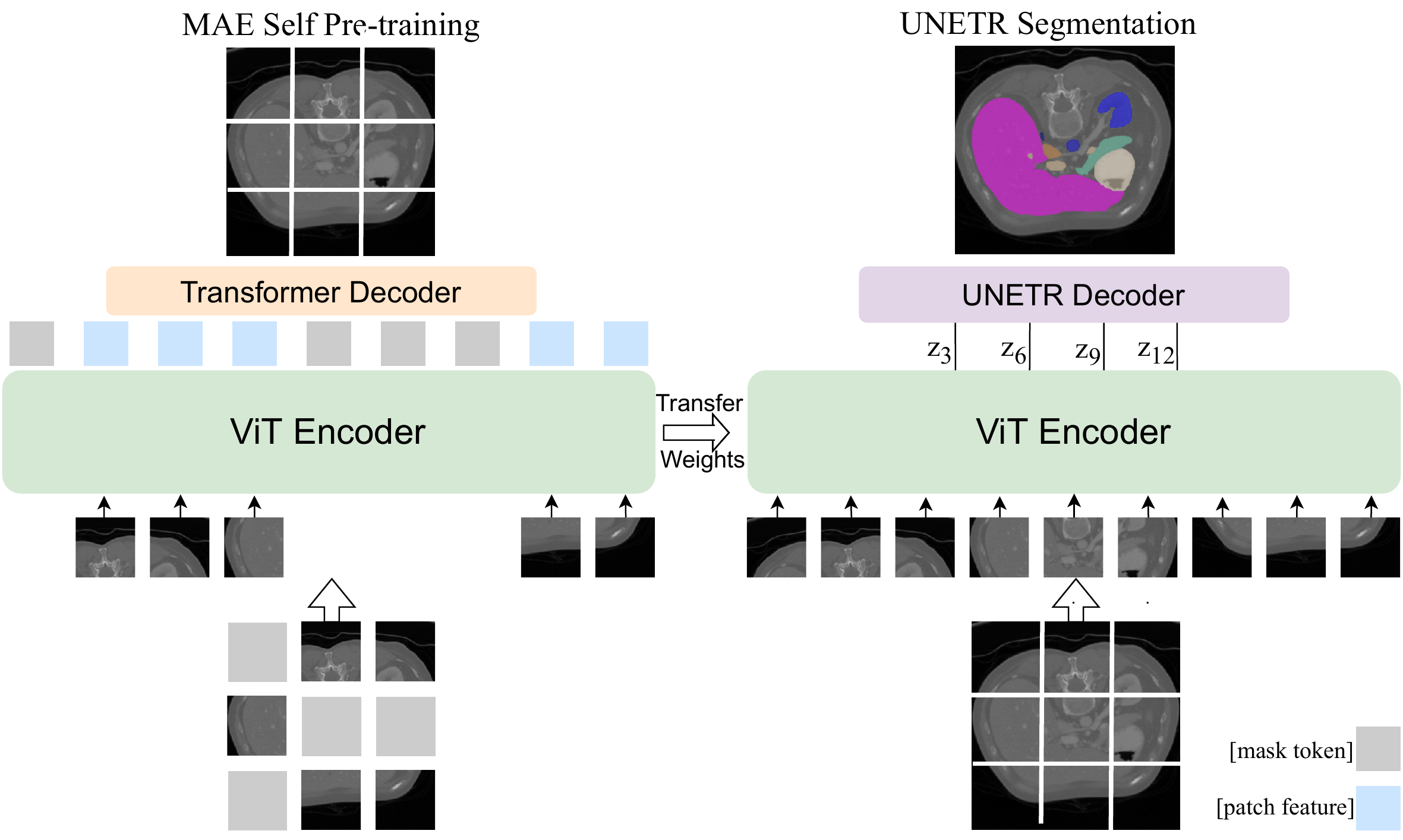}
    \caption{\textbf{Segmentation Pipeline with MAE Self Pre-training.} Left: A ViT encoder is first pre-trained with MAE. A random subset of patches is input to the encoder and a transformer decoder reconstruct the full image. Right: The pre-trained ViT weights are transferred to initialize the segmentation encoder. Then the whole segmentation network, e.g., UNETR, is finetuned to segment. A linear classifier can also be appended to an MAE self pre-trained ViT for classification tasks.}
    \label{fig:pipeline}
\end{figure}

Recent advancements in self-supervised learning (SSL) show masked image modeling (MIM)~\cite{bao2021beit,he2022masked} is an effective pre-training strategy for the Vision Transformer (ViT) \cite{dosovitskiy2020image}. The idea of MIM is \textit{masking} and \textit{reconstructing}, i.e., \textit{masking} a set of image patches at the input and \textit{reconstructing} masked patches at the output.
In this way, MIM encourages the network to \textit{infer the masked target by aggregating information from the context}. 
We believe the ability to aggregate contextual information is essential for medical image analysis. Among different MIM frameworks, Masked Autoencoder (MAE) \cite{he2022masked} is both simple and effective.
MAE has an asymmetric encoder-decoder architecture, with a ViT encoder that is input with only visible tokens, and a lightweight decoder that reconstructs the masked patches from the encoder patch-wise output and trainable mask tokens. MAE is trained with a mean square error loss by regressing the input pixel values.

In this paper, we propose a MAE-based self pre-training paradigm for medical image analysis tasks. We apply MAE pre-training on the same dataset (train-set) as the downstream task. We term this \textit{self pre-training}.
Self pre-training can benefit more scenarios where suitable pre-training data is hard to acquire. 
It also avoids the domain discrepancy between pre-training and fine-tuning by unifying the training data of two stages. We experiment on three medical image tasks including lung disease classification on chest X-ray14 (CXR14~\cite{wang2017chestx}), CT multi-organ segmentation on BTCV~\cite{landman2015miccai} and MRI brain tumor segmentation (BraTS) from the Medical Segmentation Decathlon~\cite{antonelli2021medical}. 
After MAE pre-training, ViT is added with a task-specific head and fine-tuned for downstream tasks. For classification, the head is a linear classifier. For segmentation, we follow the decoder design of UNETR~\cite{hatamizadeh2022unetr}. Note that UNETR is only a segmentation framework which does not include any self-supervised pre-training algorithm.

Our experimental results indicate that MAE self pre-training can significantly improve medical image segmentation and classification performance compared to random initialization. MAE self pre-training also surpasses the ImageNet pre-training paradigm on all the datasets.

\section{Methodology}
\subsection{Preliminary: Vision Transformer}
\label{vit}
We use ViT as the backbone for both pre-training and downstream tasks. 
A ViT is composed of a patch embedding layer, position embedding, and Transformer blocks.\\
\textbf{Patch Embedding:} The patch embedding in ViT layer needs to transform data into sequences. 3D volumes $\mathbf{x} \in \mathbb{R}^{H\times W \times D \times C}$ are first reshaped into a sequence of flattened 3D patches $\mathbf{x}_p \in \mathbb{R}^{N\times (P^3\cdot C)}$, where $C$ is the input channel, $(H, W, D)$ is the resolution, $(P, P, P)$ is the patch resolution, and $N = HWD/P^3$ is the number of patches, i.e., the length of the input sequence fed into the Transformer. A trainable linear projection is applied to map them to patch embeddings.\\
\textbf{Position Embedding:} To retain positional information, patch embeddings are added with position embeddings. The standard ViT adopts 1D learnable position embeddings. However, we experimentally find that the learnable 1D position embeddings can hurt the reconstruction of MAE. Therefore, we use sine-cosine~\cite{he2022masked,chen2021empirical} position embedding in the pre-training stage. For the downstream tasks, we initialize the learnable position embedding with sine-cosine embedding values.\\
\textbf{Transformer Block:} This consists of alternating layers of multiheaded self-attention (MSA)~\cite{vaswani2017attention} and MLP blocks.
\begin{figure}[t]
    \centering
    \includegraphics[width=0.46\textwidth]{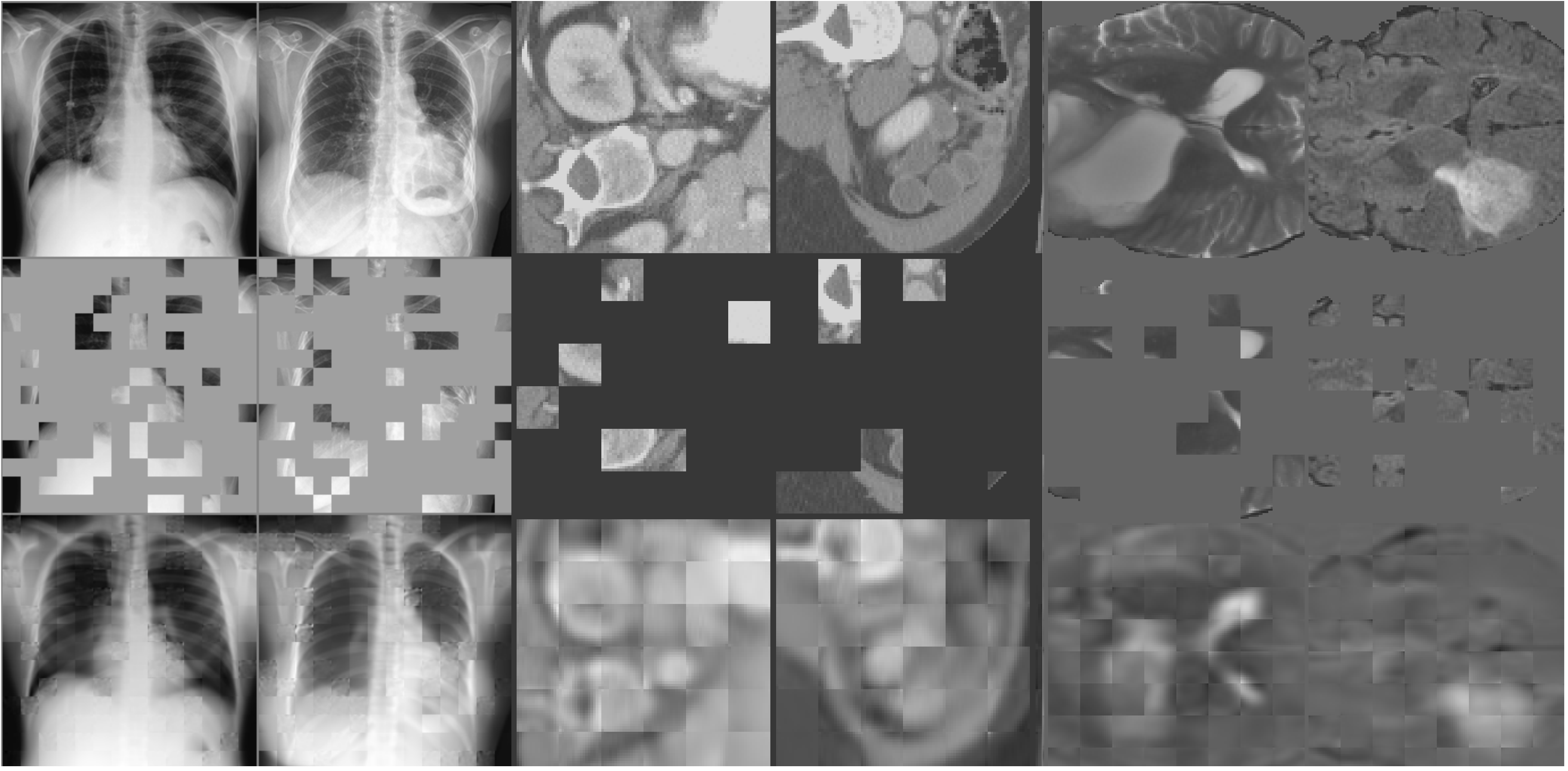}
    \caption{\textbf{MAE reconstruction.} First row: Original image. Second row: Masked image where masked regions are colored with gray/black. Third row: Reconstructed images from unmasked patches. Every two columns show the results of one dataset, i.e., CXR, BTCV and BraTS from left to right. 
    }
    \label{fig:mae_vis}
\end{figure}

\subsection{Self Pre-training with Masked Autoencoders}
\label{self-mae}
In this section, we illustrate the components in MAE, i.e., the encoder, the decoder and the loss function.\\
\textbf{Encoder in MAE.} As shown in Fig.~\ref{fig:pipeline}(Left), the ViT encoder reconstructs the full input from only partial patches. The input is first divided into non-overlapping patches. Patches are then randomly assigned into visible and masked groups.
The MAE encoder only operates on visible patches to learn a representation. To restore the positional information, patches are also added with corresponding position embeddings before being forwarded into ViT. Since the encoder output is used to reconstruct the masked input, the encoder is encouraged to extract a global representation from partial observations.\\
\textbf{Decoder in MAE.} 
The MAE decoder is input with the full set of tokens including patch-wise representations from the encoder and learnable mask tokens put in the positions of masked patches. 
By adding positional embeddings to all input tokens, the decoder can restore the patch in each specific masked position.
Note that the decoder is only an auxiliary module for pre-training and is not used in downstream tasks.\\
\textbf{Loss function in MAE.} MAE is trained with a reconstruction loss, i.e., mean squared error. Instead of reconstructing the complete image/volume, i.e., both visible and masked patches, MAE only predicts the pixel/voxel values of the masked patches, which is proven to achieve better results~\cite{he2022masked}.
In practice, normalized pixel/voxel values within each patch are better reconstruction targets than raw values.

\begin{table*}[ht]
\caption{\textbf{Abdomen Multi-organ Segmentation on BTCV.} MAE self pre-training improves upon the UNETR baseline by a large margin from 78.8\% to 83.5\% on DSC. It also shows superior performance to ImageNet supervised pre-training.}
    \centering
    \scalebox{0.7}{
    \begin{tabular}{cccccccccc}
    \hline
    Framework & Avg DSC$\uparrow$/HD95$\downarrow$ & Aorta & Gallbladder & Kidney(L) & Kidney(R) & Liver & Pancreas & Spleen & Stomach \\
    \hline
    U-Net(R50)~\cite{ronneberger2015u} & 74.68/36.87 & 84.18 & 62.84 & 79.19 & 71.29 & 93.35 & 48.23 & 84.41 & 73.92 \\
    AttnUNet(R50)~\cite{schlemper2019attention} & 75.57/36.97 & 55.92 & 63.91 & 79.20 & 72.71 & 93.56 & 49.37 & 87.19 & 74.95 \\
    TransUNet~\cite{chen2021transunet} & 77.48/31.69 & 87.23 & 63.13 & 81.87 & 77.02 & 94.08 & 55.86 & 85.08 & 75.62 \\
    \hline
    DSTUNet~\cite{cai2022dstunet} & 82.44/17.83 & 88.16 & 67.40 & 87.46 & 81.90 & 94.17 & 66.16 & \textbf{92.13} &  82.10 \\
    nnFormer~\cite{zhou2021nnformer} & 86.57/\textbf{10.63} & 92.04 & 70.17 & \textbf{86.57} & 86.25 & 96.84 & \textbf{83.35} & 90.51 & \textbf{86.83} \\
    nnUNet~\cite{isensee2021nnu} & \textbf{86.99}/10.78 & \textbf{93.01} & \textbf{71.77} & 85.57 & \textbf{88.18} & \textbf{97.23} & 83.01 & 91.86 & 85.26 \\
    \hline
    UNETR & 78.83/25.59 & 85.46 & 70.88 & 83.03 & 82.02 & 95.83 & 50.99 & 88.26 & 72.74 \\
    UNETR+ImageNet & 79.67/24.28 & 86.07 & 74.29 & 82.44 & 81.65 & 95.84 & 58.08 & 87.74 & 69.98 \\
    UNETR+MAE & \textbf{83.52}/\textbf{10.24} & \textbf{88.92} & \textbf{75.25} & \textbf{86.37} & \textbf{84.00} & \textbf{95.95} & \textbf{65.02} & \textbf{90.56} & \textbf{80.89} \\
    \hline
    \end{tabular}
    }
    \label{tab:btcv}
\end{table*}

\subsection{Architectures for Downstream Tasks}
\label{downarch}
\begin{table}[h]
\caption{\textbf{Lung Disease Classification on CXR14.} We compare MAE self pre-training with training from scratch, training with longer epochs, and training with transferred ImageNet weights. MAE self pre-training outperforms them all.}
    \centering
    \scalebox{0.7}{
    \begin{tabular}{ccccc}
        \hline
        \multirow{2}{*}{Architectures} & \multicolumn{2}{c}{Pre-training} & \multirow{2}{*}{Epochs} & \multirow{2}{*}{mAUC$\uparrow$}\\
        \cline{2-3}
        & Method & Dataset & & \\
        \hline
        CXR14-R50~\cite{wang2017chestx} & supervised & ImageNet-1K & - & 74.5\% \\
        ChestNet~\cite{wang2018chestnet} & supervised & ImageNet-1K & - & 78.1\% \\
        CheXNet~\cite{rajpurkar2017chexnet}\cite{zhou2022lung} & supervised & ImageNet-1K & - & 78.9\% \\
        ResNet18~\cite{he2016deep} & MoCo~\cite{he2020momentum} & Self & - & 78.6\% \\
        ResNet50 & MoCo-v2~\cite{chen2020improved} & Self & - & 79.4\% \\
        $Enc_t$~\cite{zhou2022lung} & LSAE & self & - & 79.0\% \\
        \hline
        ViT-B/16 & None & None & 100 & 74.4\% \\
        ViT-B/16 & None & None & 400 & 74.9\% \\
        ViT-B/16 & supervised & ImageNet-1K & 100 & 80.7\% \\
        \hline
        ViT-B/16 & MAE & Self & 100 & \textbf{81.5}\% \\
        \hline
    \end{tabular}
    }
    \label{tab:cxr14}
\end{table}
After MAE self pre-training, we append task-specific heads to perform downstream tasks such as classification and segmentation. For classification, we append a linear classifier after the class token output from ViT. Since chest X-ray can carry more than one label, we finetune the whole network with a binary cross entropy loss. For segmentation, we build UNETR~\cite{hatamizadeh2022unetr} upon the MAE pre-trained ViT encoder and a randomly-initialized convolutional decoder. UNETR has been recently proposed for 3D image segmentation tasks. Its architecture shares the same idea of U-Net~\cite{ronneberger2015u}, i.e. features from multiple resolutions of the encoder are skip-connected with the decoder. The input to the UNETR decoder is a sequence of representations
from the encoder.
Each representation is reshaped to restore the spatial dimension.
They are further upsampled and concatenated with shallower features repeatedly for higher segmentation resolution.

\section{Experiments and Results}

\begin{table*}[ht]
\caption{\textbf{MSD Brain Tumor Segmentation \& Ablation Study on Mask Ratios and Pre-training Epochs.}
}
    \centering
    \scalebox{0.58}{
    \begin{tabular}{@{\extracolsep{4pt}}ccccccccc}
    \hline
    \multirow{2}{*}{Method} & \multicolumn{2}{c}{Average} & \multicolumn{2}{c}{WT} & \multicolumn{2}{c}{ET} & \multicolumn{2}{c}{TC} \\
    \cline{2-3} \cline{4-5} \cline{6-7} \cline{8-9} 
    & DSC$\uparrow$ & HD95$\downarrow$ & DSC$\uparrow$ & HD95$\downarrow$ & DSC$\uparrow$ & HD95$\downarrow$ & DSC$\uparrow$ & HD95$\downarrow$ \\
    \hline
    UNETR & 77.40 & 7.78 & 90.25 & \textbf{6.79} & 61.45 & 8.33 & 80.51 & 7.57 \\
    UNETR+ImageNet & 77.78 & 7.38 & 90.34 & 7.19 & 62.23 & 7.86 & 80.78 & \textbf{7.00}\\
    UNETR+MAE & \textbf{78.91} & \textbf{7.22} & \textbf{90.84} & 7.04 & \textbf{63.88} & \textbf{7.15} & \textbf{82.00} & 7.13 \\
    \hline
    \end{tabular}
    \begin{tabular}{cccccc}
    \hline
    Mask ratio & \makecell{Pre-train \\ Epochs} & Avg DSC$\uparrow$ & WT & ET & TC \\
    \hline
    87.5\% & 500 & 77.14 & 90.22 & 61.06 & 80.15 \\
    \hline
    75\% & 500 & 78.14 & 90.60 & 62.48 & 81.35 \\
    75\% & 1000 & 78.29 & 90.25 & 63.06 & 81.55 \\
    75\% & 2000 & 78.43 & 90.33 & 63.45 & 81.52 \\
    \hline
    50\% & 500 & 78.42 & 90.59 & 63.05 & 81.63 \\
    25\% & 500 & 78.71 & 90.76 & 63.48 & 81.88 \\
    12.5\% & 500 & \textbf{78.91} & \textbf{90.84} & \textbf{63.88} & \textbf{82.00} \\
    \hline
    \end{tabular}
    \begin{tabular}{ccc}
    \hline
    Mask ratio & \makecell{Pre-train \\ Epochs} & Avg DSC$\uparrow$ \\
    \hline
    87.5\% & 10k & 82.21 \\
    \hline
    75\% & 10k & 82.76 \\
    75\% & 40k & 81.09 \\
    \\
    \hline
    50\% & 10k & 83.2 \\
    25\% & 10k & 83.18 \\
    12.5\% & 10k & \textbf{83.52} \\
    \hline
    \end{tabular}
    }
    \label{tab:brats}
\end{table*}

\subsection{Datasets and Implementation Details}
\noindent \textbf{Lung Disease Classification on ChestX-ray14.} ChestX-ray14~\cite{wang2017chestx} is a large-scale CXR database consisting of 112,120 frontal-view CXRs from 32,717 patients. We conduct the classification task based on the official split which consists of trainval ($\sim80\% $) and testing  ($\sim20\%$) sets.
We adopt the multi-class AUC as the performance metric.\\
\textbf{Abdomen Multi-organ Segmentation on BTCV.} BTCV~\cite{landman2015miccai} (Multi Atlas Labeling Beyond The Cranial Vault) consists of 30 subjects with abdominal CT scans where 13 organs were annotated by radiologists. Each CT volume has $85\sim 198$ slices of $512 \times 512$ pixels, with a voxel spatial resolution of ($0.54 \times 0.98 \times [2.5 \sim 5.0]$ $mm^3$). We follow~\cite{chen2021transunet, transunet-github} to split the 30 cases into 18 (training) and 12 (validation). We report the average Dice similarity coefficient (DSC) and 95\%  Hausdorff Distance (HD) on 8 abdominal organs (aorta, gallbladder, spleen, left kidney, right kidney, liver, pancreas, and stomach) to align with~\cite{chen2021transunet} for ease of comparison.\\
\textbf{Brain Tumor Segmentation on MSD.} This is one of the 10 tasks in Medical Segmentation Decathlon (MSD) Challenge~\cite{antonelli2021medical}. The entire set has 484 multi-modal (FLAIR, T1w, T1-Gd and T2w) MRI brain scans. The ground-truths of segmentation includes peritumoral edema, GD-enhancing tumor and the necrotic/non-enhancing tumor core. The performance is measured on three recombined regions, i.e., tumor core, whole tumor and enhancing tumor. We randomly split the dataset into training (80\%) and validation (20\%) sets. Reported metrics include average DSC and 95\% HD.

Our experiments are implemented on PyTorch~\cite{paszke2019pytorch} and MONAI~\cite{MONAI_Consortium_MONAI_Medical_Open_2020}. We use ViT-B/16 as the backbone and AdamW as the optimizer in all the experiments. The patch size is $16 \times 16$ for 2D images and $16\times16\times16$ for 3D volumes.\\
\textbf{Data Preprocessing and Augmentation.} For ChestX-ray14, we perform  histogram equalization on all the X-ray images. During training, we randomly flip and crop a $224 \times 224$ region out of the original $256 \times 256$ image. For BTCV, we clip the raw values between -175 and 250, and re-scale the range within [0,1]. During pre-training and fine-tuning, we randomly flip and crop a $96\times 96\times 96$ volume as the input. For BraTS, we perform an instance-wise normalization over the non-zero region per channel. In pre-training and fine-tuning, we randomly flip and crop a $128\times 128\times128$ volume.\\
\textbf{MAE Self Pre-training.} The initial learning rate (lr) is 1.5e-4 and weight decay is 0.05 for all tasks. lr decays to zero following a cosine schedule with warm-ups.
MAE pre-training runs for 800 epochs on ChestX-ray14, 10,000 on BTCV, and 500 on BraTS with training batch sizes as 256, 6 and 6.

\begin{figure}[t]
    \centering
    \includegraphics[width=.4\textwidth]{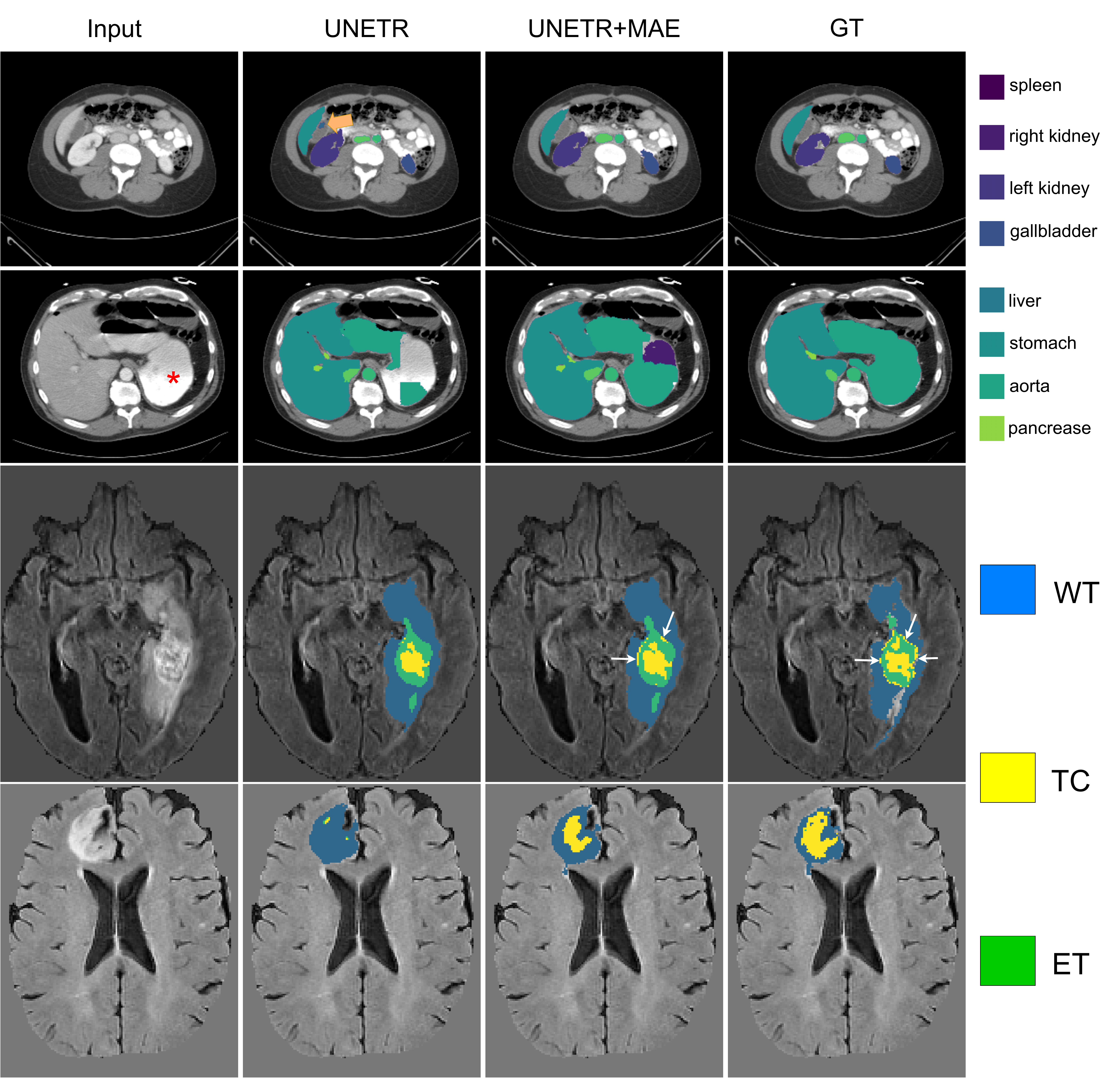}
    \caption{\textbf{Qualitative Results of Segmentation.} Results on BTCV are shown in the first two rows. In the first row, note the absence of the false positive segmentation (orange arrow) after MAE pre-training. In the second row, the stomach segmentation (red asterisk) is incomplete when created by the UNETR approach compared to an MAE pre-trained UNETR. Results on MSD BraTS are shown in the last two rows. In the third row, only subtle improvements are noticeable in the segmentation such as peripheral yellow necrotic core segmentations (white arrows) being captured after MAE pre-training. In the fourth rows, necrotic core segments are nearly absent without MAE pre-training.
    }
    \label{fig:full_seg}
\end{figure}
\noindent \textbf{Task Fine-tuning.} We adopt layer-wise learning rate decay (layer decay ratio: 0.75) to stablize the ViT training and a random DropPath with a 10\% probability. The learning rates vary between tasks. The learning rate is 1e-3 for CXR14, 8e-4 for BTCV, and 4e-4 for BraTS. The training batch size on CXR14, BTCV and BraTS is 256, 6, and 6. Learning rate during fine-tuning also follows a cosine decay schedule.

\subsection{Results}
\label{results}
\textbf{MAE Reconstruction.}
We show the reconstruction results of MAE with a mask ratio of 75\% in Fig~\ref{fig:mae_vis}. The three rows show the original images, the masked images, and the reconstructed images, respectively. The results demonstrate that MAE is able to restore the lost information from the random context. As the reconstruction loss is only applied to the masked patches, the restored visible patches look blurrier.
It is worth noting that the ultimate goal of MAE is to benefit downstream tasks rather than generating high-quality reconstructions. We hypothesize that the contextual information for reconstructions of masked image patches is of particular importance in medical imaging tasks where any given ROI is intrinsically dependent and connected to its physiological environment and surroundings.
\\
\textbf{Lung Disease Classification.}
The results are listed in Table~\ref{tab:cxr14}. First, with no pre-training, ViT hardly achieves a competitive result even with longer training epochs. This is expected considering its large model size and lack of inductive bias. Second, MAE self pre-training outperforms the ImageNet pre-trained ViT by 0.8\%. This shows the promising potential of the new MAE self pre-training paradigm for medical images. Finally, we compare with the CNN-based SOTA using both ImageNet pre-training and self-supervised pre-training by MoCo~\cite{he2020momentum,chen2020improved} and LSAE~\cite{zhou2021chest}. ViT with MAE self pre-training outperforms them all. 
\\
\textbf{Abdomen Multi-organ Segmentation.}
The results of multi-organ segmentation are shown in Table~\ref{tab:btcv}.
MAE self pre-training improves upon the UNETR baseline by a large margin, from 78.8\% to 83.5\% on average DSC. It is also superior to the ImageNet pre-training paradigm. As MAE is previously only applied on large-scale datasets like ImageNet (N=1000,000), it is interesting to observe its notable performance on a small-scale dataset (N=30); this further demonstrates the promising potential of applying MAE self pre-training to medical images in limited data scenarios. Note that the goal of our work is to demonstrate the effectiveness of self MAE pre-training instead of pursuing the state-of-the-art. With more advanced backbone architectures and data pre-processing procedures, \cite{cai2022dstunet,zhou2021nnformer,isensee2021nnu} achieve better performance than self MAE pre-trained UNETR.
\\
\textbf{Brain Tumor Segmentation.}
Results are listed in Table~\ref{tab:brats}. UNETR achieves an average DSC of 77.4\% and a HD95 of 7.78mm. With the help of MAE (12.5\% mask ratio) self pre-training, the performance of UNETR is improved further achieving a 78.91\% DSC and a HD95 of 7.22mm.
\\
\textbf{Ablation Study.}
We conduct experiments with different MAE pre-training epochs and mask ratios (Table~\ref{tab:brats}). First, the performance of MAE on BraTS generally benefits from longer training. However, prolonged pre-training can lead to inferior performance after a large number of epochs due to over-fitting. Second, unlike the high mask ratio~\cite{he2022masked} adopted in natural images, the two segmentation tasks show different preference to the mask ratios.
The best segmentation results are achieved with a mask ratio of 12.5\%.

\section{Conclusion}
We have demonstrated that MAE pre-training improves SOTA classification and segmentation performance on a diverse set of medical image analysis tasks. Importantly for medical imaging tasks, MAE self pre-training outperforms existing methods on small datasets, including ImageNet-transfer learning. Furthermore, we demonstrate the effectiveness of MAE on 3D medical images including both CTs and MRIs.
In future work, we will test the efficacy of MAE pretraining in prognosis and outcome prediction tasks~\cite{bae2021predicting}.

\section{Compliance with ethical standards}
\label{sec:ethics}

This  research  study  was  conducted  retrospectively  using open access images. Additional approval was not required.

\section{Acknowledgments}
\label{sec:acknowledgments}

The reported research was partly supported by NIH award $\#$ 1R21CA258493-01A1, NSF awards IIS-2212046 and IIS-2123920, and Stony Brook OVPR seed grants. The content is solely the responsibility of the authors and does not necessarily represent the official views of the National Institutes of Health.  The authors have no relevant financial or non-financial interests to disclose.

\bibliographystyle{IEEEbib}
\bibliography{strings,refs}

\end{document}